\documentclass[doublecol]{epl2}
\usepackage{amsmath}
\usepackage{amsfonts}
\usepackage{amssymb}
\usepackage{graphicx}
\usepackage{dcolumn}
\usepackage{bm}

\title{Cusps in the quench dynamics of a Bloch state}

\author{J.~M.~Zhang\inst{1,2} \and Hua-Tong Yang\inst{3} }

\institute{
  \inst{1} Fujian Provincial Key Laboratory of Quantum Manipulation and New Energy Materials,
College of Physics and Energy, Fujian Normal University, Fuzhou 350007, China\\
  \inst{2} Fujian Provincial Collaborative Innovation Center for Optoelectronic Semiconductors and Efficient Devices, Xiamen 361005, China \\
  \inst{3} School of Physics, Northeast Normal University, Changchun 130024, China
}

\abstract{We report some nonsmooth dynamics of a Bloch state in a one-dimensional tight binding model with the periodic boundary condition. After a sudden change of the potential of an arbitrary site, quantities like the survival probability of the particle in the initial Bloch state show cusps periodically, with the period being the Heisenberg time associated with the energy spectrum. This phenomenon is a \emph{nonperturbative} counterpart of the nonsmooth dynamics observed previously (Zhang and Haque, arXiv:1404.4280) in a periodically driven tight binding model. Underlying the cusps is an exactly solvable model, which consists of equally spaced levels extending from $-\infty$ to $+\infty $, between which two arbitrary levels are coupled to each other by the same strength. }


\begin{document}
\maketitle

\section{Introduction}

In quantum mechanics, there exist two parallel themes \cite{tanor}. One is about the static properties of a system, namely the eigenstates and eigenvalues of the Hamiltonian. The other is about the dynamics of the system, namely how the wave function or the expectation values of various physical quantities evolve. While for the former, there exist many theorems which give us a good picture of the wave functions in many cases; for the latter, the relevant mathematics is far less developed, and hence we often have little intuition. Actually, the dynamics of a system can be very surprising \cite{akulin}. This is the case even in the single-particle case, as demonstrated by the celebrated phenomena of dynamical localization \cite{dl} and coherent destruction of tunneling \cite{cdt}.


Here we report some unexpected dynamics in the setting of the one-dimensional tight binding model, which is arguably  the simplest model in solid state physics. It is about a very simple scenario. Take a tight binding model with periodic boundary condition and put a particle in some eigenstate, i.e., a Bloch state with some momentum. Then suddenly quench it by changing the potential of an arbitrary site. The rough picture is that the particle will be reflected by the newly introduced barrier, and the particle will perform Rabi oscillation between the initial Bloch state and its time-reversed counterpart. However, exact numerical simulation reveals the unexpected fact that the curves of some physical quantities like the probability of finding the particle in the initial state, are structured. Specifically, they show \emph{cusps} periodically in time.

The cusps here are somehow similar to the cusps observed previously in Ref.~\cite{scienceopen}, which are also in the tight binding model setting (the cusps there were observed earlier in quantum optics settings \cite{parker, meystre} but were not fully accounted for). The only difference in the scenario is that there the defect potential is modulated sinusoidally instead of being held fixed. However, the crucial difference is that, there the cusps (called kinks) are a perturbative effect and survive only in the weak driving limit, while here they are a generic \emph{nonperturbative} effect and thus are very \emph{robust}.

In the following, we shall first describe the phenomenon by presenting the numerical observations. Then we will identify the essential features of the underlying Hamiltonian, from which we define an idealized model. The phenomenon is then accounted for by solving the dynamics of the ideal model analytically. Finally, we discuss its physical implications.

\section{Periodically appearing  cusps}\label{phenomenon}

\begin{figure*}[tb]
\centering
\includegraphics[width= 0.99 \textwidth ]{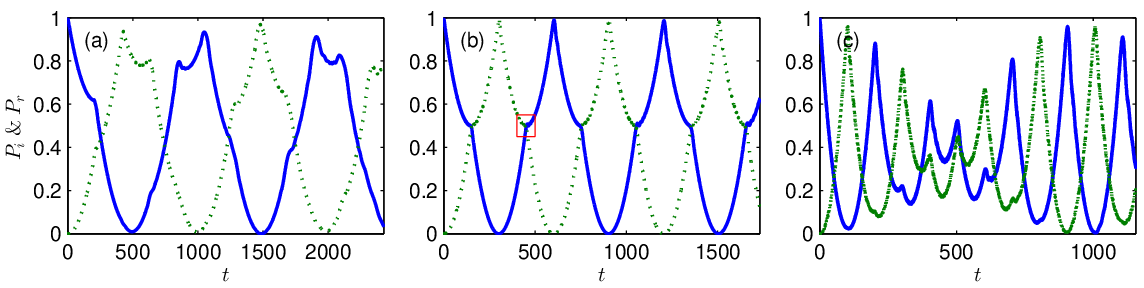}
\caption{(Color online) Time evolution of the probability of finding the particle in the initial Bloch state $|k_i\rangle $ ($P_i$, solid lines) and in the momentum-reversed Bloch state $|-k_i\rangle $ ($P_r$, dotted lines). Note that $P_i+P_r \neq 1$ in general as other Bloch states are occupied too, but $P_i +P_r =1$ to a good accuracy when the cusps show up. The values of the parameters $(N, k_i, U)$ are $(401, 80, 1.5)$, $(301, 75, 2)$, and $ (201, 50, 12)$ in (a), (b), and (c), respectively.
\label{evidence}}
\end{figure*}

\begin{figure}[tb]
\centering
\includegraphics[width= 0.40 \textwidth ]{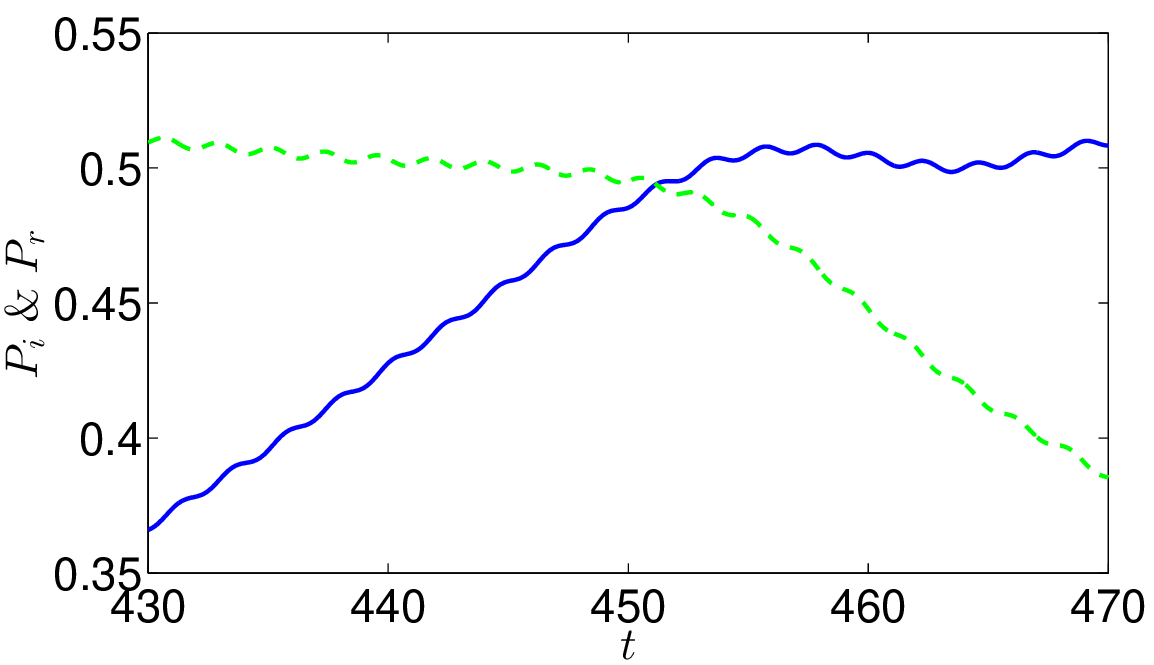}
\caption{(Color online) Details of the cusps enclosed by the red box in Fig.~\ref{evidence}(b). The cusps are smoothed on a small time scale.
\label{details}}
\end{figure}

The Hamiltonian of an $N$-site tight binding model with the periodic boundary condition is ($\hbar = 1$ throughout this paper)
\begin{equation}\label{tbmh}
  \hat{H}_0 = - \sum_{l=0}^{N-1} ( |l\rangle \langle l+1 | + |l+1 \rangle \langle l |) .
\end{equation}
Here $|l\rangle  $ is the Wannier state on site $ l $. The eigenstates are the well-known Bloch states $|k \rangle $ defined as $\langle l |k\rangle = \exp(i q l )/\sqrt{N}$.  Here $k$ is an integer defined up to an integral multiple of $ N $ and $q = 2\pi k / N$ is the associated wave vector.

Now consider such a scenario. Initially the particle is in some Bloch state $|k_i \rangle $. Then at time $t= 0 $, the potential on some site $j $ is suddenly changed to $U$. That is, we add the term $\hat{H}_1 = U |j \rangle \langle j | $ to the Hamiltonian (\ref{tbmh}). Because of the periodic boundary condition, we can assume $j  = 0 $ without loss of generality.  In the ensuing nontrivial dynamics, two quantities of particular interest are the survival probability and the reflection probability
\begin{eqnarray}
  P_i (t) = |\langle +k_i | \Psi(t) \rangle |^2,  \quad   P_r (t)= |\langle -k_i | \Psi(t) \rangle |^2,
\end{eqnarray}
which are, respectively, the probability of finding the particle in the initial Bloch state and the momentum-reversed Bloch state. Both quantities can be easily calculated  numerically as in Fig.~\ref{evidence}. There we show the numerical results of $P_i$ and $P_r$ as functions of time.

The most prominent feature of the curves is the cusps. In each panel of Fig.~\ref{evidence}, the cusps are equally spaced in time. They appear simultaneously in the curves of $P_i$ and $P_r$. Sometimes, the cusp in one of the two curves is not so clearly visible, but the corresponding one in the other curve is well shaped. Of all of the three panels, panel (b) is especially regular. Not only the cusps appear periodically, both curves are simply periodic. Moreover, when the cusps happen, $P_{i,r} = 0.5$ or $1$.

One might wonder whether the cusps are cusps in the mathematical sense as the system is a finite one. Indeed, they are not. In Fig.~\ref{details}, the cusps enclosed by the red box in Fig.~\ref{evidence}(b) are magnified---They are quite smooth. Hence, the cusps appear to be cusps only on a relatively large time scale. However, as we shall see below, behind the rounded cusps here is an exactly solvable model consisting of infinitely many levels, where the cusps are cusps in the rigorous sense.

It is worthy to emphasize the essential difference between the cusps here and those observed previously in Ref.~\cite{scienceopen}. There, it is a first order perturbative effect. The cusps exist only in the weak driving limit, or specifically, only when the survival probability $P_i$ is close to unity, and between the cusps the survival probability is a linear function of time. In contrast, here apparently the cusps are still very sharp even when $P_i$ constantly drops to zero. Moreover, the functional form of the curves between the cusps is neither linear nor exponential, but as we shall see below, \emph{quadratic}.

\section{Explanation by a truncated and linearized model}\label{solution}

\begin{figure}[tb]
\centering
\includegraphics[width= 0.4 \textwidth ]{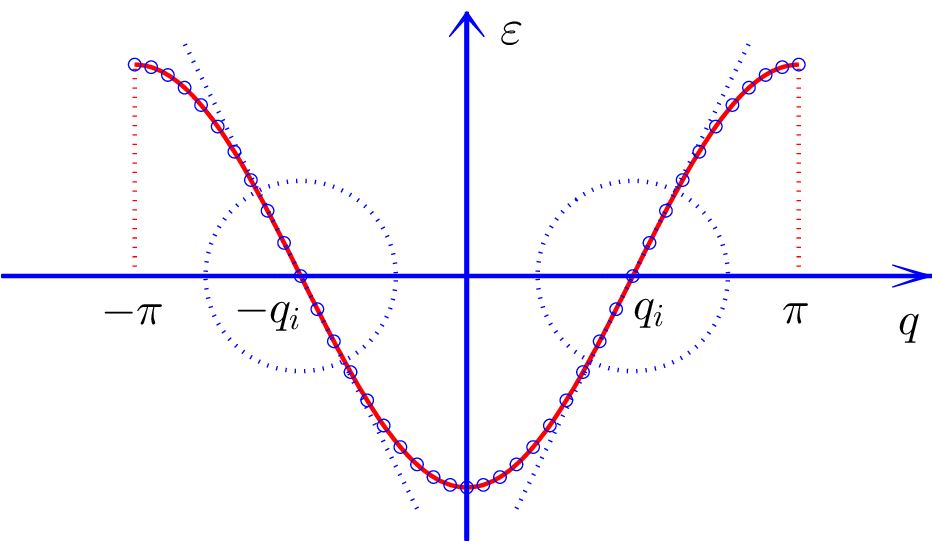}
\caption{(Color online) Dispersion relation $\varepsilon( q)= -2 \cos q $ of the tight binding model (\ref{tbmh}). The parameter $q_i = 2\pi k_i/ N$ denotes the wave vector of the initial Bloch state. The dotted straight lines are local linear approximations to the dispersion curve. Only the Bloch states inside the circles participate significantly in the dynamics and thus are retained in the truncated Hamiltonian.
\label{plot}}
\end{figure}

To account for the cusps in Fig.~\ref{evidence}, we need to have a close survey of the structures of the un-perturbed Hamiltonian $\hat{H}_0$ and the perturbation $\hat{H}_1$. Figure~\ref{plot} shows the dispersion relation, $\varepsilon (q) = -2 \cos q$, of $\hat{H}_0$. The perturbation $\hat{H}_1$ couples two arbitrary Bloch states with an equal amplitude
\begin{equation}\label{couple}
  g = \langle k_1 | \hat{H}_1 | k_2 \rangle = U/N ,
\end{equation}
regardless of the values of $k_1$ and $ k_2$.

A crucial fact revealed by numerics is that, in the evolution of the wave function, essentially only those few Bloch states with wave vectors $q \simeq \pm q_i$ contribute significantly to the wave function. Now since locally the dispersion curve $\varepsilon (q)$ can be approximated by a straight line (it is especially the case at $q = \pm \pi/2 $ where $\varepsilon''(q) = 0 $), we are led to truncate and linearize the model.

\begin{figure}[tb]
\centering
\includegraphics[width= 0.3 \textwidth ]{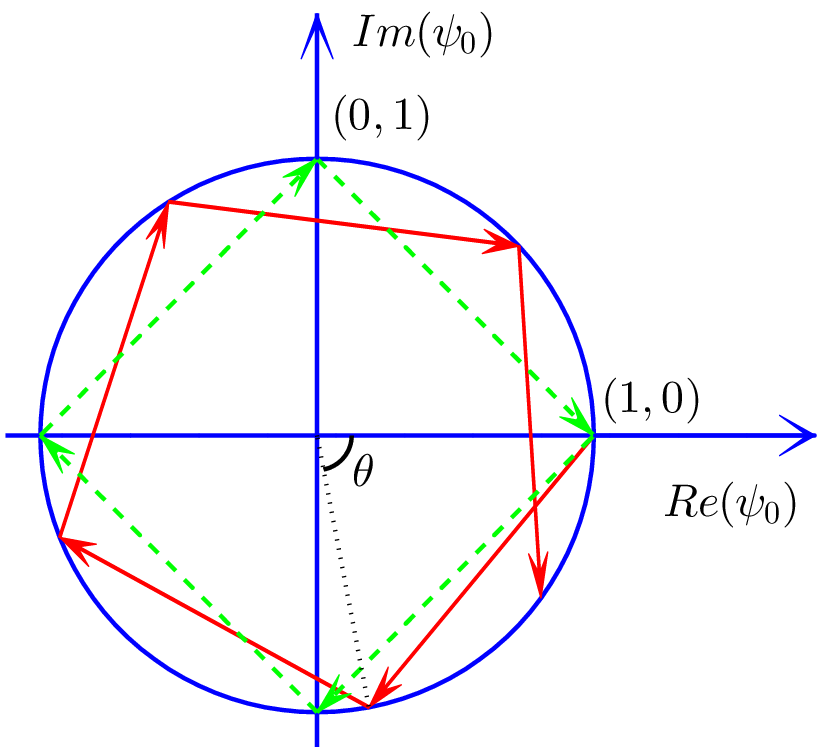}
\caption{(Color online) A generic (red solid lines) trajectory of $\psi_0 $ on the complex plane according to Eq.~(\ref{final1}a). It is analogous to the bouncing of a classical ball inside a circular billiard. The green dashed closed trajectory corresponds to the case of $\theta = \pi/2 $.
\label{trajectory}}
\end{figure}

Of all the Bloch states, we retain only two groups centered at $|\pm k_i\rangle $. Each group consists of $2M+1$ (the conditions on $M$ will be discussed later) states with wave numbers symmetrically distributed around $k_i $ or $-k_i $. Let us now refer to them as $\{| R_n \rangle \}$ and $\{ | L_n \rangle  \}$, where $R$ and $L$ mean right-going and left-going, respectively, and $n$ ranges from $-M$ to $M$. By choice, $|R_n\rangle \equiv  |k_i +n \rangle$ and $|L_n \rangle \equiv | -k_i-n \rangle$. After linearizing the dispersion curve at $\pm q_i $, the energy of the degenerate states $|R_n\rangle $ and $|L_n\rangle $ is $ n \Delta$, with
\begin{eqnarray}\label{delta}
\Delta =  ( 4 \pi \sin q_i ) /N .
\end{eqnarray}
Here we have chosen the energy of $|\pm k_i \rangle $ as the zero of energy. Again, the perturbation $\hat{H}_1$ couples two arbitrary states in the retained set of states with equal amplitude $g= U/N$.

This truncated and linearized model can be partially diagonalized by introducing a new basis as
$  |A_n^{\pm} \rangle = (|R_n \rangle \pm  |L_n\rangle )/\sqrt{2} $.
Referring to the original Hamiltonian $\hat{H}_0 $, they are even- and odd-parity states with respect to the defected site, respectively. It is easy to see that $|A_n^-\rangle $, which are eigenstates of the original Hamiltonian $\hat{H}_0$ with eigenvalues $n \Delta$, are also eigenstates of the total Hamiltonian $\hat{H } = \hat{H}_0 + \hat{H}_1 $ with the same eigenvalues. In the yet to be diagonalized subspace of $\{ |A_n^+ \rangle \}$, the matrix elements of $\hat{H}_0$ and $\hat{H}_1 $ are
\begin{eqnarray}\label{idealmodel}
 \quad  \langle A_{n_1}^+ | \hat{H}_0 | A_{n_2}^+ \rangle=  n_1 \Delta \delta_{n_1, n_2 } , \;
  \langle A_{n_1}^+ | \hat{H}_1 | A_{n_2}^+ \rangle = 2 g  ,
\end{eqnarray}
for arbitrary $n_{1,2}$.

Now the scenario is like this. Initially the system is in the state $ |\Psi(0)\rangle  =|k_i \rangle =  | R_0\rangle $. The problem is, how does the probability $P_i $ of finding the system in the initial level $| R_0\rangle $ evolve in time? We have the decomposition
$  |\Psi (0 ) \rangle =  (|A_0^- \rangle + |A_0^+ \rangle )/\sqrt{2}  $.
Since $|A_0^-\rangle $ is an eigenstate of $\hat{H}$, we see that at an arbitrary time later, the wave function has the form
\begin{equation}\label{form}
  |\Psi(t)\rangle = \frac{1}{\sqrt{2}} |A_0^- \rangle + \frac{1}{\sqrt{2}} \sum_{n=-M}^M  \psi_n (t ) |A_n^+ \rangle.
\end{equation}
By the initial condition, $\psi_n (0) = \delta_{n,0 }$. The probabilities $P_i $ and $P_r $, which are our primary concern, can be expressed as
\begin{eqnarray}\label{pipr2}
P_i =  \frac{1}{4}\left | 1+ \psi_0 \right |^2 , \quad P_r =  \frac{1}{4}\left | 1- \psi_0 \right |^2.
\end{eqnarray}
Hence, the aim is to calculate $\psi_0(t)$.

\begin{figure*}[tb]
\centering
\includegraphics[width= 0.425 \textwidth ]{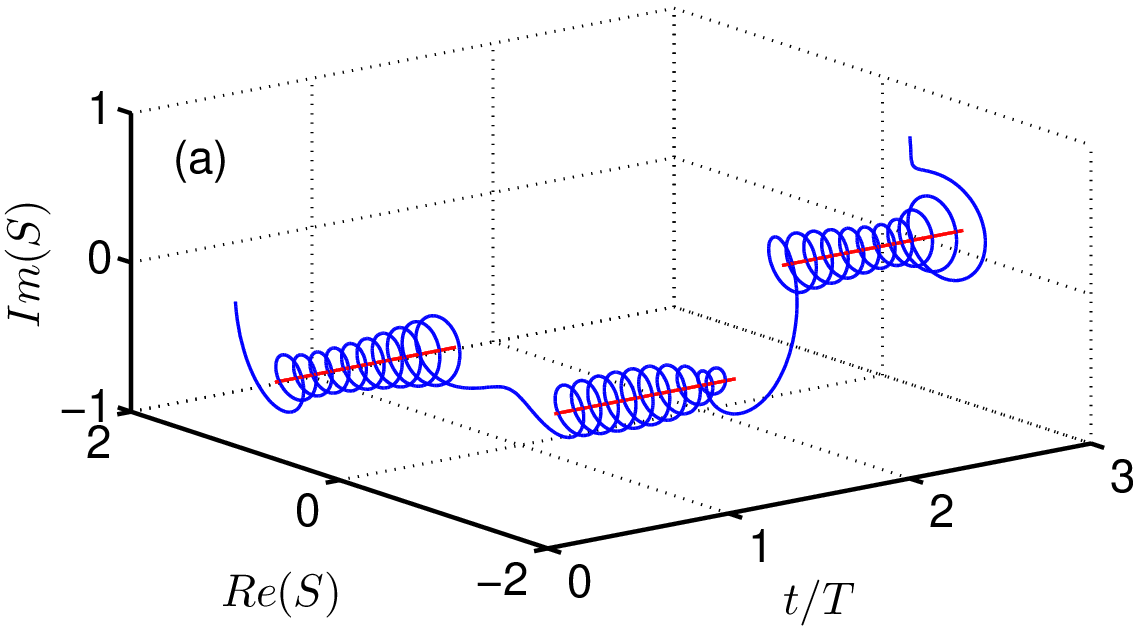}
\includegraphics[width= 0.425 \textwidth ]{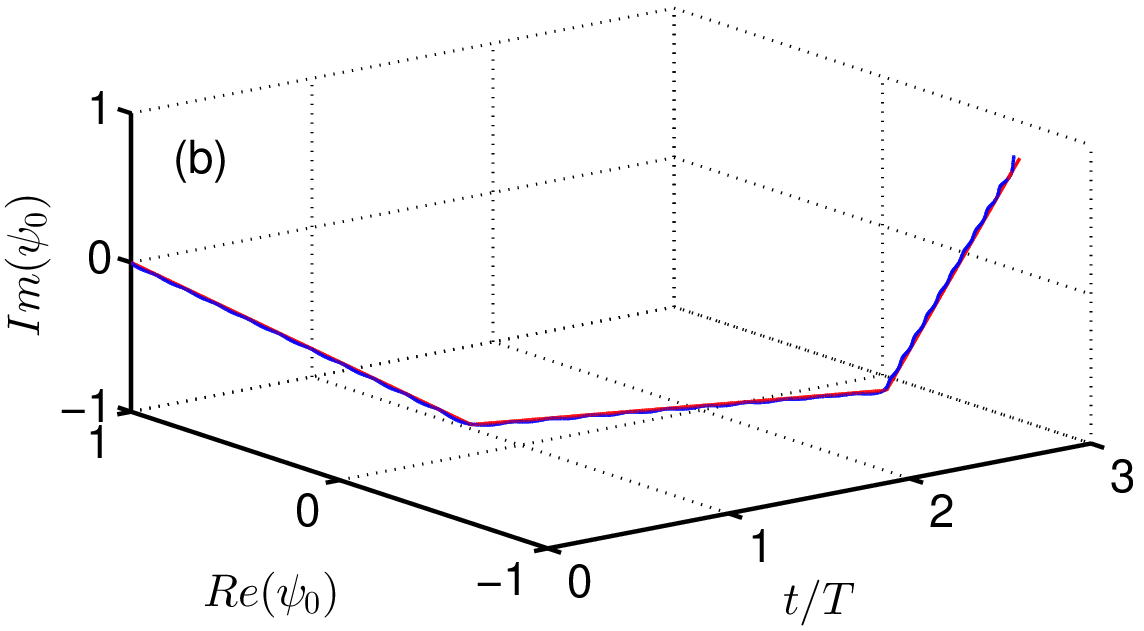}
\caption{(Color online) Time evolution (blue solid lines) of (a) the auxiliary quantity $S$ and (b) $\psi_0$ for $M = 10$ and $g/ \Delta = 0.125 $. In each panel, the red line indicates the analytical predictions of (\ref{final1}a) or (\ref{final1}b). Compare (b) with Fig.~\ref{trajectory}.
\label{trajS}}
\end{figure*}

Projecting the time-dependent Schr\"odinger equation $i \frac{\partial }{\partial t} |\Psi(t)\rangle = \hat{H} |\Psi(t)\rangle $ onto $|A_n^+ \rangle $, we get $   i \frac{\partial }{\partial t } \psi_n = n \Delta  \psi_n +  2g S  $,
where $S$ is a collective, auxiliary quantity defined as
\begin{equation}\label{S}
  S(t) \equiv  \sum_{m=-M}^{M} \psi_m (t) .
\end{equation}
The point is that it is independent of $n $. Introducing the so-called Heisenberg time $T = 2\pi /\Delta $, we see that for fixed $M$, with respect to the reduced time $t/T$, the dynamics of the truncated model is determined by the ratio $g/\Delta $ alone.

\begin{figure}[tb]
\centering
\includegraphics[width= 0.4 \textwidth ]{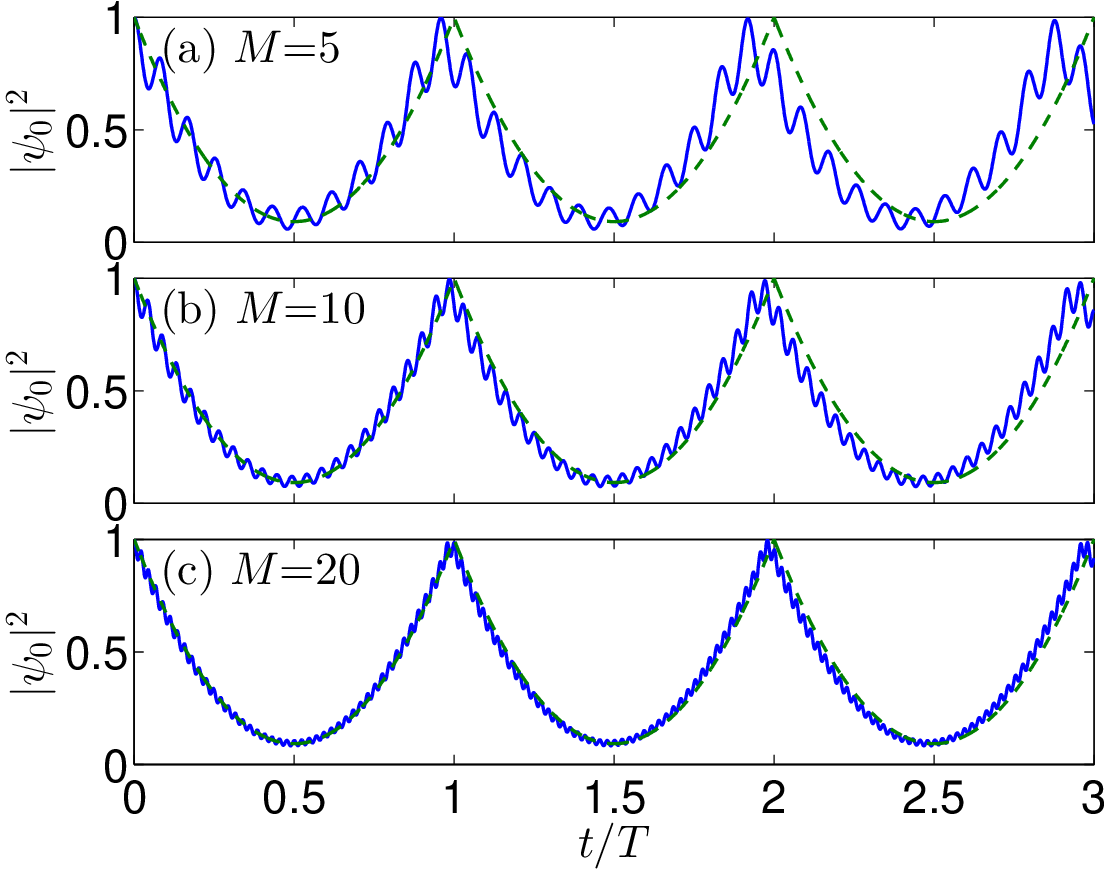}
\caption{(Color online) Time evolution of $|\psi_0|^2$ for  finite values of $M$, with $g/ \Delta =0.5$. In each panel, the blue solid line indicates the numerical exact value while the green dashed  line the analytical formula (\ref{final1}a), which corresponds to the $M\rightarrow \infty $ limit. In each period, the latter is a parabola.
\label{check}}
\end{figure}

The quantity $\psi_n $ can be solved formally as
\begin{equation}\label{solu2}
  \psi_n (t) = e^{-i n \Delta  t } \delta_{n,0} - i 2  g \int_0^t d \tau e^{-i n \Delta  (t-\tau)} S(\tau) .
\end{equation}
Plugging this into (\ref{S}), we get an integral equation of $S$,
\begin{equation}\label{S2}
  S (t) = 1  - i2 g \int_0^t d \tau \left( \sum_{n=-M}^{M } e^{-i n \Delta (t-\tau)}  \right) S(\tau) .
\end{equation}
Here to proceed analytically further, we use some fact verified by numerics (see Figs.~\ref{trajS} and \ref{check}). Numerically, it is observed that as  $M \rightarrow   \infty $, the trajectories of $\psi_n $ (in particular, $\psi_0$) converge quickly. Therefore, for sufficiently large $M $, it is legitimate to replace the finite summation in the bracket by an infinite one. That is,
\begin{equation}\label{S3}
  S (t) \simeq  1 - i 2g \int_0^t d \tau \left( \sum_{n=-\infty}^{+\infty } e^{-i n\Delta  (t-\tau)}  \right) S(\tau) .
\end{equation}
Using the Poisson summation formula \cite{grafakos}, the kernel $  \sum_{n=-\infty}^{+\infty } e^{-i n \Delta  (t-\tau)}$ can be rewritten as
$ T  \sum_{n = -\infty }^{+\infty} \delta (t - \tau - n T ) $. Substituting this new form of the kernel into (\ref{S3}), we get
$  S (t) = 1  -i g T  S(t) $,
for $0< t < T  $, by noting that $\int_0^\infty dt \delta(t) = 1/2 $. We then solve for $0< t < T  $,
\begin{equation}\label{S5}
  S(t) = \frac{1}{1 + i g T  }  ,
\end{equation}
which is a constant. Substituting this into (\ref{solu2}), we get
\begin{equation}\label{firstperiod1}
 \psi_{0}(t) = 1 - \frac{i 2g  t}{1 + i g T  } = \frac{1- i2g( t- T/2) }{1+ i g T  } ,
\end{equation}
which is linear in $t$. We note that as $t\rightarrow T^-  $,
\begin{equation}\label{lim1}
  {\psi}_{0}(t) \rightarrow \frac{1- i gT }{1+ i g T } = e^{-i\theta }
\end{equation}
for some $\theta \in (-\pi, \pi) $. That is, after one period, $\psi_0 $ returns to its initial value, except for a phase accumulated. This complete revival means, $\psi_n (t+ T) = \psi_n (t) e^{-i\theta}$ for all $n$. We thus have for $ t=  rT +s $, with $r$ being a nonnegative integer and $s \in [0, T) $,
\begin{subequations}\label{final1}
\begin{eqnarray}
  {\psi}_{0}(t) &=  & \frac{1- i2g(  s-   T/2 ) }{1+ i g T } e^{-ir\theta }, \\
  S(t) &= &\frac{1}{1+ i g T}  e^{-i r \theta } .
\end{eqnarray}
\end{subequations}
In Fig.~\ref{trajectory}, the trajectory of $\psi_0$ on the complex plane is illustrated. It bounces inside the unit circle elastically like a ball. Hence, its kinematics has the regularity of the irrational rotation \cite{irrational}. We see that $|\psi_0|^2$ is a periodic function of time $t$. At $t=r  T$, it returns to unity and in-between it is a quadratic function of $t$. By (\ref{pipr2}), $P_i+ P_r = (1+|\psi_0|^2)/2$. Hence, generally $P_i+P_r < 1 $ as $|\psi_0 |^2<1$, but at $t =r T $, when $|\psi_0|^2 =1 $, we have $P_i+P_r =1$, which is satisfied to a good accuracy in Fig.~\ref{evidence}.

\begin{figure*}[tb]
	\centering
	\includegraphics[width= 0.99 \textwidth ]{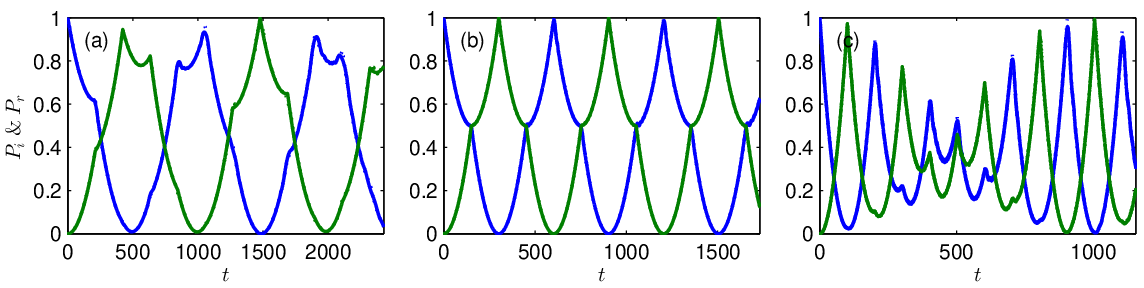}
	\caption{(Color online) Comparison between the numerical exact values of $P_{i,r}$ and the analytical predictions. The panels correspond to those in Fig.~\ref{evidence} one to one and in order. The analytical curves are solid (respectively, dotted) if the corresponding numerical curves are dotted (respectively, solid). The dotted lines are hardly visible, which proves that the numerical and analytical results agree with each other very well.
		\label{compare}}
\end{figure*}

A peculiar feature of (\ref{S5}) and (\ref{final1}b) is that $S$ is not continuous at $t= rT$. For example, by the definition (\ref{S}), $S(t=0 ) = \psi_0(t=0) = 1$, however by (\ref{S5}), $S(t=0^+ ) \neq 1$. This should be an artifact of our treatment involving the $M\rightarrow \infty $ limit. To see how this difficulty is solved for finite $M $, we demonstrate the typical time evolution behavior of $S $ with $ M =10$ in Fig.~\ref{trajS}(a). We see that in the interval of $rT < t <(r+1) T$, $S$ oscillates rapidly around the constant value predicted by (\ref{final1}b), and at about $t= rT $, the orbit of $S $ quickly transits from around one constant value to around the next. Along with the time evolution of $S$ in Fig.~\ref{trajS}(a), we show in Fig.~\ref{trajS}(b) the time evolution of $\psi_0 $. We see that the numerical exact value of $\psi_0 $ follows the analytical prediction of (\ref{final1}a) closely, with much smaller oscillation amplitude than $S $. This is reasonable in view of (\ref{solu2}), where $S$ appears in the integral and thus its oscillation is averaged out.

Further evidences demonstrating that the simple formula (\ref{final1}a) is a good approximation for finite $M$ (Anyway, there are only a finite number of levels in the original tight binding model) are presented in Fig.~\ref{check}. There we see that even for $M=5$, the formula (\ref{final1}a) captures the behavior of $|\psi_0 |^2$ on the scale of $T$ very well, and as $M$ increases, the curve converges to that predicted by (\ref{final1}a) very quickly.

Having verified that (\ref{final1}a) is reliable even for finite levels, we now apply the theory to the original problem.
There we have $\Delta = 4\pi \sin q_i /N$ and $g = U/N$. Using (\ref{pipr2}) and (\ref{final1}a), we can calculate $P_{i,r}$ in Fig.~\ref{evidence} analytically. The results are presented in Fig.~\ref{compare} together with those numerical data in Fig.~\ref{evidence}. We see that the analytical approximation and the numerical exact results agree very well.
We can also understand the regularity of Fig.~\ref{evidence}(b) now. For $U=2$ and $q_i = \pi/2$, $gT =1$ (regardless of the value of $N$) and hence $\theta = \pi/2$ and the trajectory of $\psi_0$ is the closed one in Fig.~\ref{trajectory}. By (\ref{pipr2}), it results in the regular behavior of $P_{i,r}$ in Fig.~\ref{evidence}(b).

Another regularity in all panels of Fig.~\ref{evidence} and Fig.~\ref{compare} is that, by Fig.~\ref{trajectory}, the cusps of $P_{i,r}$ are located on the curves of $(1\pm \cos \omega t)/2$ respectively, with $\omega = \theta /T  $. This fact is in accord with the rough picture that in the long term, the particle performs Rabi oscillation between the two Bloch states $| \pm k_i \rangle$. But since $\omega \neq 2 g $, we see that, because of coupling to other Bloch states, the oscillation frequency is not simply determined by the direct coupling $g$ between the two. From the point of view of quantum chaos, the system in question has a very regular dynamics.

\begin{figure}[tb]
\centering
\includegraphics[width= 0.45 \textwidth ]{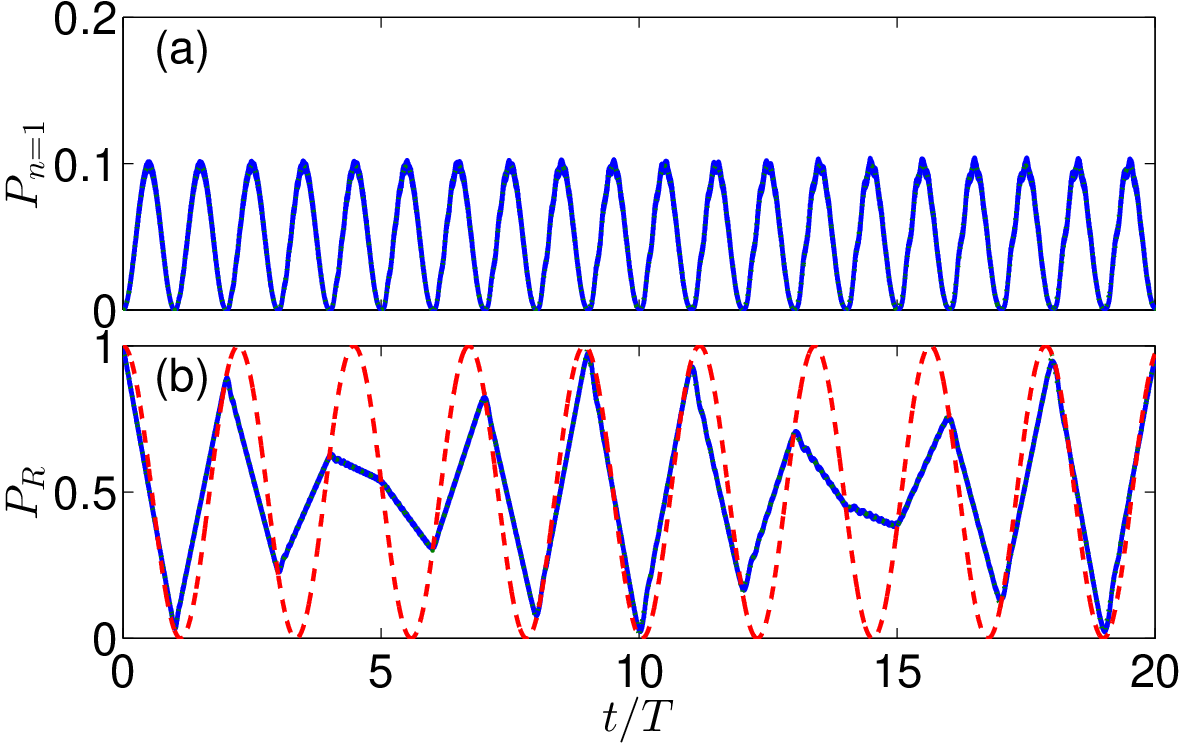}
\caption{(Color online) (a) Population on the Bloch state $|k_i + 1\rangle $ and (b) total population on those right-moving Bloch states. In each panel, there are actually two curves, the blue solid one for the numerical exact result and the green dotted one for the analytical approximation [Eqs.~(\ref{pn}) and (\ref{Pr3})]. But they coincide well. The scenario is as in Fig.~\ref{evidence}(c). In (b), the tips of the cusps are located on the red dashed curve of $(1+ \cos \omega t)/ 2$ with $\omega = \theta /T $.
\label{kinks}}
\end{figure}

Finally, we are prepared to discuss the conditions on $M $. On the one hand, $M $ should be large enough so that the $M\rightarrow \infty $ limit of the dynamics of $\psi_0 $ has almost been achieved. On the other hand, $M $ should be small enough so that the linearization approximation is valid. For fixed values of $U$ and $q_i $, as the ratio $g/\Delta = U /4\pi \sin q_i $ is independent of $N$, the lower bound set by the first condition is $N$-independent. On the contrary, the upper bound set by the second condition is obviously linearly proportional to $N $. Therefore, for arbitrary $U $ and $q_i $,  there will be room for $M $ if $N $ is sufficiently large. When it is the case, the ideal model with $M=\infty $ is a good approximation of the realistic model, as far as  the quantities $P_{i,r}$ are concerned.

\section{Populations on other Bloch states}\label{otherBS}

So far, we have focused on $\psi_0$, the quantity relevant for calculating the populations on the Bloch states $|\pm k_i\rangle $. But the exact solution above  allows us  also to calculate all other $\psi_n$, which are related to the populations on other Bloch states. By (\ref{solu2}), similar to (\ref{final1}a), we have ($t = r T + s$)
\begin{eqnarray}
  \psi_n (t) = \frac{2g }{n \Delta (1+ i g T)}(e^{-i n\Delta s} -1) e^{-ir \theta },
\end{eqnarray}
for $n \neq 0 $. By (\ref{form}), the analytic prediction of the population on the Bloch states $|\pm (k_i + n )\rangle $ is
\begin{eqnarray}\label{pn}
  P_n (t) = \frac{1}{4}|\psi_n (t )|^2 = \frac{4g^2 \sin^2 (n\Delta t/2)}{(1+g^2 T^2) n^2 \Delta^2}.
\end{eqnarray}
This is expected to be a good approximation for small $n$. Indeed, as shown in Fig.~\ref{kinks}(a), Eq.~(\ref{pn}) agrees with the numerical exact result for $P_1$ very well for the scenario in Fig.~\ref{evidence}(c). The crucial feature of (\ref{pn}) is that the amplitude of $P_n $ shrinks as $1/n^2 $. This is in line with the numerical finding that only those Bloch states in the vicinity of $|\pm k_i \rangle $ are significantly populated. It also explains why the $M = \infty $ limit is relevant for the original model, which has only a finite number of levels and a globally nonlinear spectrum---The $n$-large states are only negligibly populated both in the ideal model and in the realistic model.

Another quantity of interest is the total population on those Bloch states moving to the right, i.e.,
\begin{eqnarray}\label{Pr}
  P_R (t ) = \sum_{k=1}^{[N/2]} |\langle k | \Psi(t) \rangle |^2.
\end{eqnarray}
Here the lower and upper bounds of summation actually do not need to touch the band edges, as the contribution is primarily from the vicinity of $k_i $. By the correspondence $|k_i + n\rangle \leftrightarrow |R_n \rangle $ for small $n $, $P_R $ should can be approximated by
\begin{eqnarray}\label{Pr2}
\sum_{n\in \mathbb{Z}} |\langle R_n | \Psi (t)\rangle |^2 =\frac{1}{4} \bigg(|1+ \psi_0|^2 + \sum_{n\neq 0 } |\psi_n |^2  \bigg) .
\end{eqnarray}
Using (\ref{final1}a), (\ref{pn}), and the equality $ \sum_{n \in \mathbb{Z}} \sin^2 n \alpha/n^2 \alpha^2  =  \pi/\alpha $, for $0<\alpha < \pi$ \cite{scienceopen,fermi}, it is straightforward to reduce (\ref{Pr2}) to
\begin{eqnarray}\label{Pr3}
\cos^2 \frac{r\theta }{2 }- \sin \left[ \left(r+ \frac{1}{2}\right)\theta \right] \frac{g s }{\sqrt{1+ g^2 T^2}}.
\end{eqnarray}
It is a \emph{piecewise linear} function of $t$. As shown in Fig.~\ref{kinks}(b), it is a good approximation of the numerical exact result. The regularity here is that the tips of the cusps are located on the curve of $(1+ \cos \omega t)/2$ with $\omega = \theta / T$.

\section{Conclusions and discussions}\label{conclusion}

In conclusion, we have found the reflection dynamics of a Bloch state against a site defect to be nonsmooth, in that many quantities show cusps periodically in time. The point is that there exists an ideal model with an infinite number of levels, whose dynamics can be solved exactly and shows cusps in the mathematical sense. The model is defined by (\ref{idealmodel}), with the two characteristics of equally spaced levels and equal coupling between two arbitrary levels. The realistic tight binding model with one defected site realizes the ideal model approximately, and its dynamics is guided by that of the latter.

Admittedly, our explanation of the cusps is primarily mathematical. Physically, we note that the period of the cusps, the Heisenberg time $T = 2\pi/ \Delta $, is exactly the time for a wave packet with wave vector $ \pm q_i $ traversing the whole lattice for one loop \cite{scienceopen}. That is, the sudden jump of the slope of $P_{i,r} $ occurs when the scattered wave packet comes back to the defect site. Therefore, the phenomenon should be an interference effect.

The cusps are reminiscent of the quantum dynamical phase transition discovered by Heyl \emph{et al}. \cite{heyl}, but in a single-particle model. They are different on the one hand from those reported in Refs.~\cite{scienceopen, parker, meystre} in that they are deeply \emph{nonperturbative}, and on the other hand from those in Refs.~\cite{stey,ligare,zhou} in that the functions in-between the cusps are not exponential but \emph{quadratic} or \emph{linear}. The different behaviors stem from the different hamiltonian structures. In Refs.~\cite{stey,ligare,zhou}, the model has the level-band structure, namely, an extra level couples to a band, and there is no coupling between the levels inside the band. In contrast, here the ideal model consists of only a band, and two arbitrary levels in the band are coupled.

Although we do not believe the phenomenon reported here is universal, we do think it provides a good example demonstrating that the dynamics of a model, even the simplest one, can be very surprising. The point is that, thorough understanding of the static properties of a model (as we do for the model in question) does not imply thorough understanding of its dynamical properties. There is a big gap from the former to the latter in many cases. In view of the intensive study of nonequilibrium dynamics of many-body systems nowadays \cite{nonequilibrium}, it is worthy to emphasize that, actually the dynamics of single-body or few-body systems is already complex enough and far from being fully understood.

\textbf{Notes added in retrospect.} For quite a while, we did not understand the cusps. But now we understand more. It is a reflection of a general rule in Fourier analysis: The smoother a function is, the quicker its Fourier components decay, and vice versa. On the contrary, the more singular a function is, the more slowly its Fourier components decay, and vice versa. In a later paper [European Journal of Physics \textbf{40} (2019) 035401], we have taken a different route, a more conventional one actually, to study the dynamics of the ideal model here. We simply solve the eigenstates and eigenvalues of the ideal model and project the initial state onto this basis. The problem to calculate the survival probability is then reduced to summing up a Fourier series, which decays slowly and thus gives rise to the cusps. We have taken this point of view to study the autocorrelatio function of a particle in an infinitely deep square well potential [J. Phys. A: Math. Theor. \textbf{52} (2019) 465305]. 

Here we studied the dynamics of a single particle in the momentum space. The non-smoothness of the dynamics also shows up in the real space and in the many-body case. See EPL \textbf{116}, 10008 (2016) and PRB \textbf{97}, 075151 (2018).

\section{Acknowledgments}
%

This work is supported by the Fujian Provincial Science Foundation under grant number 2016J05004.

\end{document}